\documentclass[aps,pra,preprint,showpacs]{revtex4}

\newcommand*{\Tr}{\mathop{\mathrm{Tr}}} 
\newcommand*{\re}{\mathop{\mathrm{Re}}}

\begin{document}

\title{General expression for the quantum Zeno and anti-Zeno effects}

\author{J. Ruseckas}
\email{ruseckas@itpa.lt}
\author{B. Kaulakys}
\affiliation{Institute of Theoretical Physics and Astronomy, Vilnius
University\\ A. Go\v{s}tauto 12, 2600 Vilnius, Lithuania }
\date{\today{}}

\begin{abstract}
In this paper we investigate the quantum Zeno and anti-Zeno
effects without using any particular model of the measurement. Making a few
assumptions about the measurement process we derive an expression for the jump
probability during the measurement. From this expression the equation, obtained
by Kofman and Kurizki {[}Nature (London) \textbf{405}, 546 (2000){]} can be
derived as a special case.
\end{abstract}

\pacs{03.65.Xp, 03.65.Ta, 03.65.Yz, 42.50.Lc}

\maketitle

\section{\label{sec:intro}Introduction}

The description of the measurement process has been a problem since early
development of quantum mechanics \cite{vonNeumann}. During recent years the
measurement problem attracted much attention due to the advancement in
experimental techniques. Nevertheless, the full understanding of
quantum-mechanical measurements has not been achieved as yet. Typically, the
measurement in quantum mechanics is described by von Neumann's state reduction
(or projection) postulate \cite{vonNeumann}. However, this postulate refers only
to an ideal measurement, which is instantaneous and arbitrarily accurate. Real
measurements are represented by the projection postulate only roughly.

The so-called ``quantum Zeno effect'' is directly related to the measurement
problem. In quantum mechanics the short-time behavior of nondecay probability of
an unstable particle is not exponential but quadratic \cite{Khalfin}. The
deviation from the exponential decay has been observed by Wilkinson \textit{et
  al.} \cite{Wilkinson}. Using the behavior of nondecay probability Misra and
Sudarshan \cite{Misra} in 1977 showed, that the frequent observations can slow
down the decay. An unstable particle would never decay when continuously
observed. Misra and Sudarshan have called this effect the quantum Zeno paradox
or effect. The very first analysis does not take into account the actual
mechanism of the measurement process involved, but it is based on an alternating
sequence of unitary evolution and a collapse of the wave function. The quantum
Zeno effect has been experimentally proved \cite{Itano} in a repeatedly measured
two-level system undergoing Rabi oscillations. The outcome of this experiment
has also been explained without the collapse hypothesis
\cite{Petrosky,Frerichs,Pascazio}.

Later it was realized that the repeated measurements could not only slow down
the quantum dynamics but the quantum process may be accelerated by frequent
measurements, as well
\cite{Kofman2,Kaulakys1,Kofman1,Lewenstein,Facchi,Ruseckas1,Kofman3}. This
effect was called a quantum anti-Zeno effect. Quantum Zeno and anti-Zeno effect
were experimentally observed in an atomic tunneling process \cite{fisher}.

Simple interpretation of quantum Zeno and anti-Zeno effects was given in
Ref.~\cite{Kofman1}. Using projection postulate the universal formula describing
both quantum Zeno and anti-Zeno effects was obtained. According to
Ref.~\cite{Kofman1}, the decay rate is determined by the convolution of two
functions: the measurement-induced spectral broadening and the spectrum of the
reservoir to which the decaying state is coupled.

In this paper we analyze the quantum Zeno and anti-Zeno effects without using
any particular measurement model and making only few assumptions. We obtain a
more general expression for the jump probability during the measurement.
Expression, derived in Ref.~\cite{Kofman1} is a special case of our formula.

The work is organized as follows. In Sec. \ref{sec:mod} we present the
description of the measurement. A simple case is considered in Sec.
\ref{sec:meas-unpert}. In Sec. \ref{sec:pert} we derived a general formula for
the probability of the jump into another level during the measurement. The
pulsed measurements when there is a period of the measurement-free evolution
between the measurements is analyzed in Sec. \ref{sec:free}. Particular case of
the expression, obtained in Sec. \ref{sec:pert}, is investigated in Sec.
\ref{sec:simpl}. Section \ref{sec:concl} summarizes our findings.

\section{\label{sec:mod}Description of the measurement}

We consider a system that consists of two parts. The first part of the system
has the discrete energy spectrum. The Hamiltonian of this part is $\hat{H}_0$.
The other part of the system is represented by Hamiltonian $\hat{H}_1$.
Hamiltonian $\hat{H}_1$ commutes with $\hat{H}_0$. In a particular case the
second part can be absent and $\hat{H}_1$ can be zero. The operator $\hat{V}(t)$
causes the jumps between different energy levels of $\hat{H}_0$. Therefore, the
full Hamiltonian of the system is of the form
$\hat{H}_S=\hat{H}_0+\hat{H}_1+\hat{V}(t)$. The example of such a system is an
atom with the Hamiltonian $\hat{H}_0$ interacting with the electromagnetic
field, represented by $\hat{H}_1$, while the interaction between the atom and
the field is $\hat{V}(t)$.

We will measure in which eigenstate of the Hamiltonian $\hat{H}_0$ the system
is. The measurement is performed by coupling the system with the detector. The
full Hamiltonian of the system and the detector equals to 
\begin{equation}
\hat{H}=\hat{H}_S+\hat{H}_D+\hat{H}_I,\label{eq:ham}
\end{equation}
 where $\hat{H}_D$ is the Hamiltonian of the detector and $\hat{H}_I$ represents
the interaction between the detector and the measured system, described by the
Hamiltonian $\hat{H}_0$. We can choose the basis
$|n\alpha\rangle=|n\rangle\otimes|\alpha\rangle$ common for the operators
$\hat{H}_0$ and $\hat{H}_1$, 
\begin{eqnarray}
\hat{H}_0|n\rangle & = & E_n|n\rangle,\\
\hat{H}_1|\alpha\rangle & = & E_{\alpha}|\alpha\rangle,
\end{eqnarray}
 where $n$ numbers the eigenvalues of the Hamiltonian $\hat{H}_0$ and $\alpha$
represents the remaining quantum numbers.

The initial density matrix of the system is $\hat{\rho}_S(0)$. The initial
density matrix of the detector is $\hat{\rho}_D(0)$. Before the measurement the
measured system and the detector are uncorrelated, therefore, the full density
matrix of the measured system and the detector is
$\hat{\rho}(0)=\hat{\rho}_S(0)\otimes\hat{\rho}_D(0)$. The duration of the
measurement is $\tau$.

When the interaction of the detector with the environment is taken into account,
the evolution of the measured system and the detector cannot be described by a
unitary operator. More general description of the evolution, allowing to include
the interaction with the environment, can be given using the superoperators.
Therefore, we will assume that the evolution of the measured system and the
detector is given by the superoperator $\mathcal{S}(t)$. The explicit form of
the superoperator $\mathcal{S}(t)$ can be obtained from a concrete model of the
measurement.

Due to the finite duration of the measurement it is impossible to realize the
infinitely frequent measurements. The highest frequency of the measurements is
achieved when the measurements are performed one after another without the
period of the measurement-free evolution between two successive measurements.
Therefore, we model a continuous measurement by the subsequent measurements of
the finite duration and finite accuracy. After $N$ measurements the full density
matrix of the measured system and the detector is
\begin{equation}
\hat{\rho}(N\tau)=\mathcal{S}(\tau)^N\hat{\rho}(0).
\end{equation}

We assume that the density matrix of the detector, $\hat{\rho}_{D}(0)$,
is the same before each measurement. Such an assumption is valid when
the initial condition for the detector, modified by the measurement,
is restored at the beginning of each measurement or each measurement
is performed with a new detector. For example, the detector can be
an atom which is excited during the measurement. After the interaction
of the atom with the measured system is interrupted, the atom returns
to the ground state due to spontaneous emission, and the result of
the measurement is encoded in the emitted photon. Thus the initial
state of the detector is restored.

\section{\label{sec:meas-unpert}Measurement of the unperturbed system}

In this section we investigate the measurement of the unperturbed system, i.e.,
the case when $V(t)=0.$

We assume that the measurement of the unperturbed system is a quantum
non-demolition measurement
\cite{qnd-braginsky-1980,qnd-caves,qnd-unruh,qnd-braginsky-1996}. The
measurement of the unperturbed system does not change the state of the measured
system when initially the system is in an eigenstate of the Hamiltonian
$\hat{H}_0$. After such an assumption, the most general form of the action of
the superoperator $\mathcal{S}(\tau)$ can be written as
\begin{equation}
\mathcal{S}(\tau)[|n\alpha\rangle\langle m\alpha'|\otimes\hat{
\rho}_D(0)]=|n\alpha\rangle\langle m\alpha'|e^{i\omega_{
m\alpha',n\alpha}\tau}\otimes\mathcal{S}_{n\alpha,m\alpha'}(\tau)\hat{
\rho}_D(0),\label{eq:assum}
\end{equation}
where
\begin{equation}
\omega_{m\alpha',n\alpha}=\frac{1}{\hbar}(E_m+E_{\alpha'}-E_n-E_{\alpha})
\end{equation}
and the superoperator $\mathcal{S}_{n\alpha,m\alpha'}(\tau)$ acts only on the
density matrix of the detector. The full density matrix of the detector and the
measured system after the measurement is
\begin{equation}
\hat{\rho}(\tau)=\mathcal{S}(\tau)\hat{\rho}(0)=\sum_{
n\alpha,m\alpha'}|n\alpha\rangle(\rho_S)_{n\alpha,m\alpha'}e^{i\omega_{
m\alpha',n\alpha}\tau}\langle m\alpha'|\otimes\mathcal{S}_{
n\alpha,m\alpha'}(\tau)\hat{\rho}_D(0).\label{eq:full}
\end{equation}
From Eq.~(\ref{eq:full}) it follows that the non-diagonal matrix elements of the
density matrix of the system after the measurement
$(\rho_S)_{n\alpha,m\alpha'}(\tau)$ are multiplied by the quantity
\begin{equation}
F_{n\alpha,m\alpha'}(\tau)\equiv\Tr\{\mathcal{S}_{n\alpha,m\alpha'}(\tau)\hat{
\rho}_D(0)\}.\label{eq:F}
\end{equation}
 Since after the measurement the non-diagonal matrix elements of the density
matrix of the measured system should become small (they must vanish in the case
of an ideal measurement), $F_{n\alpha,m\alpha'}(\tau)$ must be also small when
$n\neq m$.

\section{\label{sec:pert}Measurement of the perturbed system}

The operator $\hat{V}(t)$ represents the perturbation of the unperturbed
Hamiltonian $\hat{H}_0+\hat{H}_1$. We will take into account the influence of
the operator $\hat{V}(t)$ by the perturbation method, assuming that the strength
of the interaction between the system and detector is large and the duration of
the measurement $\tau$ is short. Similar method was used in
Ref.~\cite{Ruseckas2}.

We assume that the Markovian approximation is valid, i.e., the evolution of the
measured system and the detector depends only on their state at the present
time. Then the superoperator $\mathcal{S}$, describing the evolution of the
measured system and the detector, obeys the equation
\begin{equation}
\frac{\partial}{\partial t}\mathcal{S}=\mathcal{L}(t)\mathcal{S},\label{eq:sup}
\end{equation}
 where $\mathcal{L}$ is the Liouvilian. There is a small perturbation of the
measured system, given by the operator $\hat{V}$. We can write
$\mathcal{L}=\mathcal{L}_0+\mathcal{L}_V$, where $\mathcal{L}_V$ is a small
perturbation. We expand the superoperator $\mathcal{S}$ into powers of $V$
\begin{equation}
\mathcal{S}=\mathcal{S}^{(0)}+\mathcal{S}^{(1)}+\mathcal{S}^{(2)}+\cdots
\label{eq:series}
\end{equation}
Then from Eq.~(\ref{eq:sup}) it follows
\begin{eqnarray}
\frac{\partial}{\partial t}\mathcal{S}^{(0)} & = &\mathcal{L}_0(t)\mathcal{S}^{
(0)},\label{eq:sup0}\\
\frac{\partial}{\partial t}\mathcal{S}^{(i)} & = &\mathcal{L}_0(t)\mathcal{S}^{
(i)}+\mathcal{L}_V(t)\mathcal{S}^{(i-1)}.\label{eq:supi}
\end{eqnarray}
We will denote as $\mathcal{S}^{(0)}(t,t_0)$ the solution of Eq.~(\ref{eq:sup0})
with the initial condition $\mathcal{S}^{(0)}(t=t_0,t_0)=1$. The formal
solutions of Eqs.~(\ref{eq:sup0}) and (\ref{eq:supi}) are
\begin{equation}
\mathcal{S}^{(0)}(t,t_0)=T\exp\left(\int_{t_0}^t\mathcal{L}_0(t')dt'\right)
\end{equation}
 and
\begin{equation}
\mathcal{S}^{(i)}(t,0)=\int_0^tdt_1\mathcal{S}^{(0)}(t,t_1)\mathcal{
L}_V(t_1)\mathcal{S}^{(i-1)}(t_1,0).\label{eq:sol1}
\end{equation}
Here $T$ represents the time-ordering. In the second-order approximation we have
\begin{eqnarray}
\mathcal{S}(t,0) & = &\mathcal{S}^{(0)}(t,0)+\int_0^tdt_1\mathcal{S}^{
(0)}(t,t_1)\mathcal{L}_V(t_1)\mathcal{S}^{(0)}(t_1,0)\nonumber\\
 &  & +\int_0^tdt_1\int_0^{t_1}dt_2\mathcal{S}^{(0)}(t,t_1)\mathcal{
L}_V(t_1)\mathcal{S}^{(0)}(t_1,t_2)\mathcal{L}_V(t_2)\mathcal{S}^{
(0)}(t_2,0).\label{eq:sup2}
\end{eqnarray}
Using Eq.~(\ref{eq:series}), the full density matrix of the measured system and
the detector can be represented as 
\begin{equation}
\hat{\rho}(t)=\hat{\rho}^{(0)}(t)+\hat{\rho}^{(1)}(t)+\hat{\rho}^{(2)}(t)
+\cdots\,,
\end{equation}
 where 
\begin{equation}
\hat{\rho}^{(i)}(t)=\mathcal{S}^{(i)}(t,0)\hat{\rho}(0).
\end{equation}

Let the initial density matrix of the system and detector is 
\begin{equation}
\hat{\rho}(0)=|i\alpha\rangle\langle i\alpha|\otimes\hat{\rho}_D(0).
\label{eq:initial}
\end{equation}
 The probability of the jump from the level $|i\alpha\rangle$ into the level
$|f\alpha'\rangle$ during the measurement is 
\begin{equation}
W(i\alpha\rightarrow f\alpha')=\Tr\{|f\alpha'\rangle\langle f\alpha'|\hat{
\rho}(\tau)\}.\label{eq:prob1}
\end{equation}
 Using the equation (\ref{eq:assum}) we can write
\begin{equation}
\mathcal{S}^{(0)}(t,t_0)\left[|n\alpha\rangle\langle m\alpha'|\otimes\hat{
\rho}_D(0)\right]=|n\alpha\rangle\langle m\alpha'|e^{i\omega_{
m\alpha',n\alpha}t}\otimes\mathcal{S}_{n\alpha,m\alpha'}^{(0)}(t,t_0)\hat{
\rho}_D(0).\label{eq:assumpt2}
\end{equation}
 From Eq.~(\ref{eq:assumpt2}) it follows that the superoperator
$\mathcal{S}_{m\alpha,m\alpha}^{(0)}$ with the equal indices does not change the
trace of the density matrix $\hat{\rho}_D$, since the trace of the full density
matrix of the measured system and the detector must remain unchanged during the
evolution.

When the system is perturbed by the operator $\hat{V}(t)$ then the superoperator
$\mathcal{L}_V$ is defined by the equation
\begin{equation}
\mathcal{L}_V(t)\hat{\rho}=\frac{1}{i\hbar}[\hat{V}(t),\hat{\rho}].\label{eq:lv}
\end{equation}
 The first-order term is
$\hat{\rho}^{(1)}(t)=\mathcal{S}^{(1)}(t,0)\hat{\rho}(0)$. Using
Eqs.~(\ref{eq:sol1}), (\ref{eq:initial}), (\ref{eq:assumpt2}), and
(\ref{eq:lv}), this term can be written as
\begin{eqnarray}
\hat{\rho}^{(1)}(t) & = &\sum_{p\alpha_1}\frac{1}{
i\hbar}\int_0^tdt_2\left(|p\alpha_1\rangle V_{p\alpha_1,i\alpha}(t_2)e^{
i\omega_{i\alpha,p\alpha_1}(t-t_2)}\langle i\alpha|\otimes\mathcal{S}_{
p\alpha_1,i\alpha}^{(0)}(t,t_2)\right.\nonumber\\
 &  & -\left.|i\alpha\rangle V_{i\alpha,p\alpha_1}(t_2)e^{i\omega_{
p\alpha_1,i\alpha}(t-t_2)}\langle p\alpha_1|\otimes\mathcal{S}_{
i\alpha,p\alpha_1}^{(0)}(t,t_2)\right)\mathcal{S}_{i\alpha,i\alpha}^{
(0)}(t_2,0)\hat{\rho}_D(0).\label{eq:rho1}
\end{eqnarray}
When $i\neq f$ then the first-order term does not contribute to the jump
probability, since from Eqs.~(\ref{eq:prob1}) and (\ref{eq:rho1}) it follows
that the expression for this contribution contains the scalar product $\langle
f\alpha'|i\alpha\rangle=0$.

For the second-order term
$\hat{\rho}^{(2)}(t)=\mathcal{S}^{(2)}(t,0)\hat{\rho}(0)$, using
Eqs.~(\ref{eq:sol1}) and (\ref{eq:assumpt2}), we obtain the equality
\begin{equation}
\Tr\{|f\alpha'\rangle\langle f\alpha'|\hat{\rho}^{(2)}(t)\}=\frac{1}{
i\hbar}\int_0^tdt_1\Tr\left\{\langle f\alpha'|\hat{V}(t_1)\hat{\rho}^{
(1)}(t_1)|f\alpha'\rangle-\langle f\alpha'|\hat{\rho}^{(1)}(t_1)\hat{
V}(t_1)|f\alpha'\rangle\right\}.\label{eq:tmp1}
\end{equation}
 In Eq.~(\ref{eq:tmp1}) the superoperator
$\mathcal{S}_{f\alpha',f\alpha'}^{(0)}$ is omitted, since it does not change the
trace. Then from Eqs.~(\ref{eq:rho1}) and (\ref{eq:tmp1}) we obtain the jump
probability
\begin{eqnarray}
W(i\alpha\rightarrow f\alpha') & = &\frac{1}{\hbar^2}\int_0^{\tau}dt_1\int_0^{
t_1}dt_2\Tr\left\{\left(V_{f\alpha',i\alpha}(t_1)V_{
i\alpha,f\alpha'}(t_2)\mathcal{S}_{i\alpha,f\alpha'}^{(0)}(t_1,t_2)e^{i\omega_{
f\alpha',i\alpha}(t_1-t_2)}\right.\right.\nonumber\\
 &  & +\left.V_{f\alpha',i\alpha}(t_2)V_{i\alpha,f\alpha'}(t_1)\mathcal{S}_{
f\alpha',i\alpha}^{(0)}(t_1,t_2)e^{i\omega_{i\alpha,f\alpha'}(t_1
-t_2)}\right)\nonumber\\
 &  &\times\left.\mathcal{S}_{i\alpha,i\alpha}^{(0)}(t_2,0)\hat{
\rho}_D(0)\right\}.\label{eq:result1}
\end{eqnarray}
Equation (\ref{eq:result1}) allows us to calculate the jump probability during
the measurement when the evolution of the measured unperturbed system is known.
The explicit form of the superoperator $\mathcal{S}_{n\alpha,m\alpha'}^{(0)}$
can be obtained from a concrete model of the measurement. The main assumptions,
used in the derivation of Eq.~(\ref{eq:result1}), are Eqs.~(\ref{eq:assum}) and
(\ref{eq:sup}), i.e., the assumptions that the quantum measurement of the
unperturbed system is non-demolition measurement and that the Markovian
approximation is valid. Thus, Eq.~(\ref{eq:result1}) is quite general.

The probability that the measured system remains in the initial state
$|i\alpha\rangle$ is
\begin{equation}
W(i\alpha)=1-\sum_{f,\alpha'}W(i\alpha\rightarrow f\alpha').
\end{equation}
After $N$ measurements the probability that the measured system remains in the
initial state equals to
\begin{equation}
W(i\alpha)^N\approx\exp(-RN\tau),
\end{equation}
where $R$ is the jump rate
\begin{equation}
R=\sum_{f,\alpha'}\frac{1}{\tau}W(i\alpha\rightarrow f\alpha').
\label{eq:27}
\end{equation}

\section{\label{sec:free}Free evolution and measurements}

In practice, it is impossible to perform the measurements one after another
without the period of the measurement-free evolution between two successive
measurements. Such intervals of the measurement-free evolution were also present
in the experiments demonstrating the quantum Zeno effect
\cite{Itano,balzer,fisher}. Therefore, it is important to consider such
measurements. This problem for the definite model was investigated in
Ref.~\cite{Ruseckas3}.

We have the repeated measurements separated by the free evolution of the
measured system. For the purpose of the description of such measurements we can
use Eq.~(\ref{eq:result1}), obtained in Sec.~\ref{sec:pert}. The duration of the
free evolution is $\tau_F$, the duration of the free evolution and the
measurement together is $\tau$. The superoperator of the free evolution without
the perturbation $\hat{V}$ is $\mathcal{S}_F^{(0)}(t)$, the superoperator of the
measurement is $\mathcal{S}_M^{(0)}(t,t_0)$. We will assume that during the
measurement the superoperator $\mathcal{L}_0$ does not depend on time $t$. Then
the superoperator $\mathcal{S}_M^{(0)}(t,t_0)$ depends only on the time
difference $t-t_0$. Therefore, we will write $\mathcal{S}_M^{(0)}(t-t_0)$
instead of $\mathcal{S}_M^{(0)}(t,t_0)$. When the free evolution comes first and
then the measurement is performed, the full superoperator equals to
\begin{equation}
\mathcal{S}_{n\alpha,m\alpha'}^{(0)}(t,t_1)=\left\{
\begin{array}{cl}
\mathcal{S}_{M\, n\alpha,m\alpha'}^{(0)}(t-t_1), &\tau>t_1>\tau_F\textrm{ and
 }\tau>t>t_1,\\
\mathcal{S}_F^{(0)}(t-t_1), &\tau_F>t_1>0\textrm{ and }\tau_F>t>t_1,\\
\mathcal{S}_{M\, n\alpha,m\alpha'}^{(0)}(t-\tau_F)\mathcal{S}_F^{(0)}(\tau_F
-t_1), &\tau_F>t_1>0\textrm{ and }\tau>t>\tau_F.\end{array}\right.
\label{eq:freemeas}
\end{equation}
Equation (\ref{eq:freemeas}) can be written as
\begin{eqnarray}
\mathcal{S}_{n\alpha,m\alpha'}^{(0)}(t,t_1) & = &\mathcal{S}_{M\,
 n\alpha,m\alpha'}^{(0)}(t-t_1)\Theta(t_1-\tau_F)+\mathcal{S}_F^{(0)}(t
-t_1)\Theta(\tau_F-t)\nonumber\\
 &  & +\mathcal{S}_{M\, n\alpha,m\alpha'}^{(0)}(t-\tau_F)\mathcal{S}_F^{
(0)}(\tau_F-t_1)\Theta(t-\tau_F)\Theta(\tau_F-t_1),\label{eq:freemeas2}
\end{eqnarray}
where $\Theta$ is Heaviside unit step function. From Eqs.~(\ref{eq:result1}) and
(\ref{eq:freemeas2}) it follows that the jump probability consists of three
terms
\begin{equation}
W(i\alpha\rightarrow f\alpha')=W_M(i\alpha\rightarrow f\alpha')
+W_F(i\alpha\rightarrow f\alpha')+W_I(i\alpha\rightarrow f\alpha'),\label{eq:29}
\end{equation}
where the jump probability during the free evolution is
\begin{equation}
W_F(i\alpha\rightarrow f\alpha')=\frac{1}{\hbar^2}\int_0^{\tau_F}dt_1\int_0^{
\tau_F}dt_2V_{f\alpha',i\alpha}(t_1)V_{i\alpha,f\alpha'}(t_2)e^{i\omega_{
f\alpha',i\alpha}(t_1-t_2)},\label{eq:30}
\end{equation}
the jump probability during the measurement
\begin{eqnarray}
W_M(i\alpha\rightarrow f\alpha') & = &\frac{1}{\hbar^2}\int_{\tau_F}^{
\tau}dt_1\int_{\tau_F}^{t_1}dt_2\Tr\left\{\left(V_{f\alpha',i\alpha}(t_1)V_{
i\alpha,f\alpha'}(t_2)\mathcal{S}_{M\, i\alpha,f\alpha'}^{(0)}(t_1-t_2)e^{
i\omega_{f\alpha',i\alpha}(t_1-t_2)}\right.\right.\nonumber\\
 &  & +\left.V_{f\alpha',i\alpha}(t_2)V_{i\alpha,f\alpha'}(t_1)\mathcal{S}_{M\,
 f\alpha',i\alpha}^{(0)}(t_1-t_2)e^{i\omega_{i\alpha,f\alpha'}(t_1
-t_2)}\right)\nonumber\\
 &  &\times\left.\mathcal{S}_{M\, i\alpha,i\alpha}^{(0)}(t_2-\tau_F)\mathcal{
S}_F^{(0)}(\tau_F)\hat{\rho}_D(0)\right\},\label{eq:31}
\end{eqnarray}
and the interference term is
\begin{eqnarray}
W_I(i\alpha\rightarrow f\alpha') & = &\frac{1}{\hbar^2}\int_{\tau_F}^{
\tau}dt_1\int_0^{\tau_F}dt_2\Tr\left\{\left(V_{f\alpha',i\alpha}(t_1)V_{
i\alpha,f\alpha'}(t_2)\mathcal{S}_{M\, i\alpha,f\alpha'}^{(0)}(t_1-\tau_F)e^{
i\omega_{f\alpha',i\alpha}(t_1-t_2)}\right.\right.\nonumber\\
 &  & +\left.\left.V_{f\alpha',i\alpha}(t_2)V_{i\alpha,f\alpha'}(t_1)\mathcal{
S}_{M\, f\alpha',i\alpha}^{(0)}(t_1-\tau_F)e^{i\omega_{i\alpha,f\alpha'}(t_1
-t_2)}\right)\mathcal{S}_F^{(0)}(\tau_F)\hat{\rho}_D(0)\right\}.\label{eq:32}
\end{eqnarray}

If we assume that the free evolution does not change the density matrix of the
detector and the perturbation $\hat{V}$ does not depend on time, we have the
jump probability during the measurement-free evolution
\begin{equation}
W_F(i\alpha\rightarrow f\alpha')=|V_{i\alpha,f\alpha'}|^2\frac{
4\sin^2\left(\frac{1}{2}\omega_{f\alpha',i\alpha}\tau_F\right)}{\hbar^2\omega_{
f\alpha',i\alpha}^2},
\end{equation}
the jump probability during the measurement
\begin{eqnarray}
W_M(i\alpha\rightarrow f\alpha') & = &\frac{1}{\hbar^2}|V_{
i\alpha,f\alpha'}|^2\int_{\tau_F}^{\tau}dt_1\int_{\tau_F}^{t_1}dt_2\Tr\left\{
\left(\mathcal{S}_{M\, i\alpha,f\alpha'}^{(0)}(t_1-t_2)e^{i\omega_{
f\alpha',i\alpha}(t_1-t_2)}\right.\right.\nonumber\\
 &  & +\left.\mathcal{S}_{M\, f\alpha',i\alpha}^{(0)}(t_1-t_2)e^{i\omega_{
i\alpha,f\alpha'}(t_1-t_2)}\right)\nonumber\\
 &  &\times\left.\mathcal{S}_{M\, i\alpha,i\alpha}^{(0)}(t_2-\tau_F)\hat{
\rho}_D(0)\right\},
\end{eqnarray}
and the interference term
\begin{eqnarray}
W_I(i\alpha\rightarrow f\alpha') & = & |V_{i\alpha,f\alpha'}|^2\frac{
2\sin\left(\frac{1}{2}\omega_{f\alpha',i\alpha}\tau_F\right)}{\hbar^2\omega_{
f\alpha',i\alpha}}\nonumber\\
 &  &\times\int_{\tau_F}^{\tau}dt_1\Tr\left\{\left(\mathcal{S}_{M\,
 i\alpha,f\alpha'}^{(0)}(t_1-\tau_F)e^{i\omega_{f\alpha',i\alpha}\left(t_1
-\frac{1}{2}\tau_F\right)}\right.\right.\nonumber\\
 &  & +\left.\left.\mathcal{S}_{M\, f\alpha',i\alpha}^{(0)}(t_1-\tau_F)e^{
i\omega_{i\alpha,f\alpha'}\left(t_1-\frac{1}{2}\tau_F\right)}\right)\hat{
\rho}_D(0)\right\}.
\end{eqnarray}

\section{\label{sec:simpl}Simplification of the expression for the jump
probability}

The expression for the jump probability during the measurement can be simplified
if the operator $\hat{V}$ does not depend on time $t$. Then
Eq.~(\ref{eq:result1}) can be written as
\begin{equation}
W(i\alpha\rightarrow f\alpha')=\frac{2}{\hbar^2}|V_{
i\alpha,f\alpha'}|^2\re\int_0^{\tau}dt_1\int_0^{t_1}dt_2e^{i\omega_{
f\alpha',i\alpha}(t_1-t_2)}\Tr\{\mathcal{S}_{i\alpha,f\alpha'}^{
(0)}(t_1,t_2)\mathcal{S}_{i\alpha,i\alpha}^{(0)}(t_2,0)\hat{\rho}_D(0)\}.
\label{eq:simpl1}
\end{equation}
Introducing the function
\begin{equation}
G(\omega)_{f\alpha',i\alpha}=|V_{i\alpha,f\alpha'}|^2\delta\left(\frac{1}{
\hbar}(E_{\alpha'}-E_{\alpha})-\omega\right)
\end{equation}
we can rewrite Eq.~(\ref{eq:simpl1}) in the form 
\begin{equation}
W(i\alpha\rightarrow f\alpha')=\frac{2\pi\tau}{\hbar^2}\int_{-\infty}^{
\infty}G(\omega)_{f\alpha',i\alpha}P(\omega)_{i\alpha,f\alpha'}d\omega,
\label{eq:result}
\end{equation}
where
\begin{equation}
P(\omega)_{i\alpha,f\alpha'}=\frac{1}{\pi\tau}\re\int_0^{\tau}dt_1\int_0^{
t_1}dt_2e^{i(\omega-\omega_{if})(t_1-t_2)}\Tr\{\mathcal{S}_{i\alpha,f\alpha'}^{
(0)}(t_1,t_2)\mathcal{S}_{i\alpha,i\alpha}^{(0)}(t_2,0)\hat{\rho}_D(0)\}.
\label{eq:spectr}
\end{equation}
Equation (\ref{eq:result}) is similar to that obtained by Kofman and Kurizki in
Ref.~\cite{Kofman1}.

Further simplification can be achieved when the superoperator $\mathcal{L}_0$
does not depend on time $t$ and the order of the superoperators in the
expression
$\Tr\left\{\mathcal{S}_{i\alpha,f\alpha'}^{(0)}(t_1,t_2)\mathcal{S}_{
i\alpha,i\alpha}^{(0)}(t_2)\hat{\rho}_D(0)\right\}$ can be changed. Under such
assumptions we have
\begin{equation}
\Tr\left\{\mathcal{S}_{i\alpha,f\alpha'}^{(0)}(t_1,t_2)\mathcal{S}_{
i\alpha,i\alpha}^{(0)}(t_2)\hat{\rho}_D(0)\right\}=\Tr\left\{\mathcal{S}_{
i\alpha,i\alpha}^{(0)}(t_2)\mathcal{S}_{i\alpha,f\alpha'}^{(0)}(t_1,t_2)\hat{
\rho}_D(0)\right\}=F_{i\alpha,f\alpha'}(t_1-t_2),\label{eq:assumpt3}
\end{equation}
where $F_{i\alpha,f\alpha'}(t)$ is defined by Eq.~(\ref{eq:F}). After changing
the variables into $u=t_1-t_2$ and $v=t_1+t_2$ from Eq.~(\ref{eq:spectr}) we
obtain
\begin{equation}
P(\omega)_{i\alpha,f\alpha'}=\frac{1}{\pi}\re\int_0^{\tau}\left(1-\frac{u}{
\tau}\right)F_{i\alpha,f\alpha'}(u)\exp\left(i(\omega-\omega_{if})u\right)du.
\label{eq:42}
\end{equation}

\subsection{Decaying system}

We consider a decaying system with the Hamiltonian $\hat{H}_{0}$
that due to the interaction with the field decays from the level $|i\rangle$
into the level $|f\rangle$. The field initially is in the vacuum state
$|\alpha=0\rangle$. Only the energy levels of the decaying system are measured
and the detector does not interact with the field. Then
$\mathcal{S}_{i\alpha,f\alpha'}^{(0)}$
and $P(\omega)_{i\alpha,f\alpha'}$ do not depend on $\alpha$ and
$\alpha'$. Using Eqs.~(\ref{eq:27}) and (\ref{eq:result}) we obtain
the decay rate of the measured system
\begin{equation}
R=\sum_{\alpha}\frac{1}{\tau}W(i0\rightarrow f\alpha)=
\frac{2\pi}{\hbar^{2}}\int_{-\infty}^{\infty}G(\omega)_{f,i}P(\omega)_{i,f}d\omega,
\end{equation}
where
\begin{equation}
G(\omega)_{f,i}=\sum_{\alpha}G(\omega)_{f\alpha,i0}.
\end{equation}
The function $P(\omega)_{i,f}$ is related to the measurement-induced
broadening of the spectral line \cite{Kofman1,Kofman3,Ruseckas1}.
For example, when instantaneous ideal measurements are performed at
time intervals $\tau$, we can take $F_{i\alpha,f\alpha'}(t)=\Theta(\tau-t)$,
where $\Theta(t)$ is the unit step function. Then from Eq.~(\ref{eq:42})
we get
\[
P(\omega)_{i,f}=\frac{2\sin^{2}\left(\frac{1}{2}
\tau(\omega-\omega_{if})\right)}{\pi\tau(\omega-\omega_{if})^{2}}.
\]
We have that the width of the function $P(\omega)_{i,f}$ increases
when the duration of the measurement $\tau$ decreases.

The equation (\ref{eq:result}) represents a universal result: the
decay rate of the frequently measured decaying system is determined
by the overlap of the reservoir coupling spectrum $G(\omega)_{f,i}$
and the measurement-modified level width $P(\omega)_{i,f}$.

Depending on the reservoir spectrum $G(\omega)_{f,i}$ and the frequency
of the measurements $1/\tau$ the inhibition or acceleration of the
decay can be obtained. If the frequency of measurements is small and,
consequently, the measurement-induced broadening of the spectral line
is much smaller than the width of the reservoir coupling spectrum,
the decay rate equals the decay rate of the unmeasured system, given
by the Fermi's Golden Rule. In the intermediate region, when the width
of the spectral line is rather small compared with the distance between
$\omega_{if}$ and the nearest maximum in the reservoir spectrum,
the decay rate grows with increase of the frequency of the measurements.
This results in the anti-Zeno effect.

If the width of the spectral line is much greater compared both with
the width of the reservoir spectrum and the distance between $\omega_{if}$
and the centrum of the reservoir spectrum, the decay rate decreases
when the frequency of measurements increases. This results in the
quantum Zeno effect.

\section{\label{sec:concl}Conclusions}

We analyze the quantum Zeno and quantum anti-Zeno effects without using any
particular model of the measurement. The general expression (\ref{eq:result1})
for the jump probability during the measurement is derived. The main
assumptions, used in the derivation of Eq.~(\ref{eq:result1}), are assumptions
that the quantum measurement is non-demolition measurement
(Eq.~(\ref{eq:assum})) and the Markovian approximation for the quantum dynamics
is valid (Eq.~(\ref{eq:sup})). We have shown that Eq.~(\ref{eq:result1}) is also
suitable for the description of the pulsed measurements, when there are
intervals of the measurement-free evolution between successive measurements
(Eqs.~(\ref{eq:29})--(\ref{eq:32})). When the operator $\hat{V}$ inducing the
jumps from one state to another does not depend on time Eq.~(\ref{eq:result}),
which is of the form obtained by Kofman and Kurizki \cite{Kofman1}, is derived
as a special case.


\end{document}